# Attention Please: Consider Mockito when Evaluating Newly Proposed Automated Program Repair Techniques

Shangwen Wang
National University of Defense Technology, Changsha, China
wangshangwen13@nudt.edu.cn

Ming Wen
The Hong Kong University of Science and Technology, Hong Kong, China
mwenaa@cse.ust.hk

Xiaoguang Mao, and Deheng Yang
National University of Defense Technology, Changsha, China
{xgmao, yangdeheng13}@nudt.edu.cn

## ABSTRACT
Automated program repair (APR) has attracted widespread attention in recent years with substantial techniques being proposed. Meanwhile, a number of benchmarks have been established for evaluating the performances of APR techniques, among which Defects4J is one of the most wildly used benchmark. However, bugs in Mockito, a project augmented in a later-version of Defects4J, do not receive much attention by recent researches. In this paper, we aim at investigating the necessity of considering Mockito bugs when evaluating APR techniques. Our findings show that: 1) Mockito bugs are not more complex for repairing compared with bugs from other projects; 2) the bugs repaired by the state-of-the-art tools share the same repair patterns compared with those patterns required to repair Mockito bugs; however, 3) the state-of-the-art tools perform poorly on Mockito bugs (Nopol can only correctly fix one bug while SimFix and CapGen cannot fix any bug in Mockito even if all the buggy locations have been exposed). We conclude from these results that existing APR techniques may be overfitting to their evaluated subjects and we should consider Mockito, or even more bugs from other projects, when evaluating newly proposed APR techniques. We further find out a unique repair action required to repair Mockito bugs named *external package addition*. Importing the external packages from the test code associated with the source code is feasible for enlarging the search space and this action can be augmented with existing repair actions to advance existing techniques.

## CCS Concepts
• **Software and its engineering** → **Error handling and recovery; Software reliability; Software testing and debugging**

## Keywords
Automated Program Repair; Defects4J; Mockito

## 1. INTRODUCTION
Automated program repair (APR) techniques aim at reducing the excessively high cost in fixing bugs [1] and have shown to be promising in increasing the effectiveness of automated debugging [2]. Pioneered by GenProg [3], substantial techniques have been proposed recently [4 – 9] and plenty of empirical studies have been conducted to assess APR towards diverse aspects [10 – 14].

To facilitate controlled experiments and fair evaluations for different APR techniques, researchers have built several publicly available benchmarks during recent years [15 – 19]. Defects4J [20], one of the most widely used benchmark, is a database containing Java bugs from real-world open source projects. Each bug in this benchmark is extracted from the project's associated version control histories through three steps, including identifying real bugs fixed by developers, reproducing those real bugs, and isolating them. As a result, each bug corresponds to two versions, buggy and fixed, and is accompanied by a comprehensive programmer-written test suite that can reveal the bug (at least one test triggers a failure on the buggy version). Due to the test execution framework provided by Defects4J can facilitate the experiments of fault localization and program repair, this database has been widely used by recent studies.

In the first release of Defects4J in June 2015, it contained 357 bugs from five open source projects, namely JFreechart, Closure compiler, Apache commons-lang, Apache commons-math, and Joda-Time, respectively. Subsequently, in October 2016, the original team added 38 new bugs from another project named Mockito into this database when releasing its version 1.1.0[1]. Many APR techniques were proposed after that [21 – 31], however, none of them have evaluated the performance on these 38 bugs. This phenomenon motivates our study. Our intuition is that since these 38 bugs are extracted by the original methodology and integrated into the database by the professional developers, they must be able to make supplementary for this database from some aspects. It seems unreasonable to exclude them from the evaluation criteria of newly proposed APR techniques.

In this paper, we conduct an empirical study to investigate the importance of taking bugs from Mockito into consideration when evaluating APR techniques. Our study can be summarized in four phases: In the first phase, we statistic the characteristics of human-written patches from Mockito project and compare them with that of other projects. Our aim is to investigate whether the patches from Mockito are more complex than patches from the other projects and the result is negative. In the second phase, we examine whether patches from Mockito contain any unique repair patterns that cannot be observed among the patches from other projects and the result is still negative. In the third phase, we choose two recently proposed techniques to evaluate their performances on Mockito and adopt the experimental results from a study which evaluated another tool on Mockito. Results indicate that the state-of-the-art APR tools perform poorly when evaluating on Mockito. To summarize: since bugs in Mockito are not more complex than bugs from other projects and they do not require extra repair patterns, the poor performances of the state-of-the-art APR tools on this project drive us to conclude that **we do need to take Mockito into consideration when evaluating newly released APR techniques**. There may exist overfitting between existing APR techniques and their evaluated subjects. Repair action is a kind of modification on source code like adding a method call that is made to fix bugs [32]. Based on this, repair patterns are abstractions occurring recurrently in patches that can involve compositions of repair actions [33]. Due to the poor performances of the state-of-the-art APR techniques on Mockito bugs, a finer-grained granularity analysis about the properties of patches in Mockito bugs is needed. Consequently, in the last phase, we further manually check all the 38 patches in Mockito to seek is there any unique repair action in these patches and we find out one named *external package addition*. We also find an instance from the other projects with the same repair action that is successfully fixed by ACS. By investigating this experience, we point out a way for searching for fixing ingredients in a larger

---
[1] https://github.com/rjust/defects4j/tree/v1.1.0

space: importing the external packages from the test code associated with the source code.

In summary, this paper makes the following contributions:
- A comprehensive comparison between Mockito and other projects in Defects4J towards the patch characteristics and repair patterns;
- An experiment on the evaluation of the performances of SimFix and CapGen, two of the latest search-based APR tools, on bugs from Mockito;
- The identification of a unique repair action in Mockito and an experience for enlarging the search space.

The reminder of this paper is organized as follows. Section 2 introduces the background of our study. Section 3 describes the research questions of this empirical study. Section 4 presents the answers to our research questions with results and analysis. Section 5 and 6 discuss the limitations and related work. Finally, Section 7 concludes the paper.

## 2. BACKGROUND

Defects4J, which represents for a database of existing faults to enable controlled testing studies for Java programs, is a wildly used database in software testing and analysis in recent years. Each real bug included in Defects4J fulfills the following three requirements. First, the bug is related to source code which means bug fixes within the build system, configuration files, documentation, or tests are not included. Second, the bug is reproducible which means at least one test that passes on the fixed version fails on the buggy version. Third, the bug is isolated, indicating that unrelated changes such as features or refactoring are not included. The developers of Defects4J also features a database abstraction layer that eases the use of the bug database and a test execution framework that provides several components for common tasks in software testing such as test execution, test generation, and code coverage or mutation analysis. All the convenient designs mentioned above make this benchmark popular in this field.

After its version 1.1.0 released in October 2016, Defects4J contains totally 395 bugs from six open source projects. Table 1 characterizes the bugs from this database. In Table 2, we list nine APR techniques designed for Java (i.e., ACS [21], ssFix [22], JAID [23], SOFix [24], ProbabilisticModel [25], CapGen [26], SketchFix [27], Elixir [28] and SimFix [29], respectively) which are proposed after this milestone with the projects they select for evaluating their performances. There are still some other techniques designed for Java being proposed after this milestone such as JFix [30] and NPEFix [31]. We do not discuss them in this section since they have not been evaluated on Defects4J.

From the results, none of these techniques select Mockito as the evaluation criteria. Four approaches (i.e., ACS, SOFix, CapGen, and Elixir) do not select Closure and three of them (except for Elixir) clearly explain the reason is that this project uses a customized testing format instead of JUnit test. However, none of them provide a clear explanation for excluding Mockito from the testing set. This newly augmented project does not receive much attention.

Our study is motivated by this observation. In our mind, since bugs from Mockito are added by the original researchers using the original methodology, they should be an important part of this database since its release. The scientific evidence for ignoring these bugs when evaluating the performance of APR technique is not strong enough. Besides, taking this project into consideration can lead to a more comprehensive verification set, eliminating the bias caused by the evaluation process. In the rest of this paper, we are dedicated to awakening researchers' attention to Mockito through our empirical study. Since our study will compare Mockito with other projects in Defects4J database frequently, we use terms Mockito and non-Mockito to distinguish them.

**Table 1. Details of the Defects4J benchmark**

| Projects | Bugs | kLoC | Tests |
|---|---|---|---|
| JFreechart (Chart) | 26 | 96 | 2,205 |
| Closure compiler (Closure) | 133 | 90 | 7,927 |
| Apache commons-math (Math) | 106 | 85 | 3,602 |
| Apache commons-lang (Lang) | 65 | 22 | 2,245 |
| Joda-Time (Time) | 27 | 28 | 4,130 |
| Mockito (Mockito) | 38 | 45 | 1,457 |
| Total | 395 | 366 | 21,566 |

In the table, column "Bugs" denotes the total number of bugs from this project in Defects4J benchmark, column "kLoC" denotes the number of thousands of lines of code, and column "Tests" denotes the total number of test cases of each project. "Tests" data is excerpted from previous study [13].

**Table 2. Details of some recently released APR tools**

| Tools | Source | Selected projects |
|---|---|---|
| ACS | ICSE'17 | Chart, Math, Lang, Time |
| ssFix | ASE'17 | Chart, Closure, Math, Lang, Time |
| JAID | ASE'17 | Chart, Closure, Math, Lang, Time |
| SOFix | SANER'18 | Chart, Math, Lang, Time |
| Probabilistic-Model | SANER'18 | Chart, Closure, Math, Lang, Time |
| CapGen | ICSE'18 | Chart, Math, Lang, Time |
| SketchFix | ICSE'18 | Chart, Closure, Math, Lang, Time |
| Elixir | ICSE'18 | Chart, Math, Lang, Time |
| SimFix | ISSTA'18 | Chart, Closure, Math, Lang, Time |

In the table, column "Source" denotes the conference the corresponding APR tool publishes in. The contents of this column are divided into two parts: the previous section represents the short name of this well-known international conference while the latter section represents the year. Note that these tools are arranged according to the order of their publication time.

## 3. RESEARCH QUESTIONS

In order to figure out our observed phenomenon clearly, we propose the following four research questions:

**RQ1: Are patches of Mockito bugs more complex than that of non-Mockito bugs in Defects4J?** In order to find out if the bugs in Mockito are more difficult to fix, this problem analyzes these patches from a quantitative perspective. To answer this question, we select four features of the patches which can help to quantify the complexity and difficulty to fix a bug (i.e., patch size, number of chunks, number of modified files, and number of modified methods). We observe the distribution situations of these attributes of bugs from each project and perform a significant difference test.

**RQ2: Are state-of-the-art techniques capable to fix non-Mockito bugs whose repairs require the same repair patterns as Mockito bugs?** This question, analyzing the repair patterns of Mockito bugs, is an extension of the previous one. To answer this question, we classify bugs in Mockito into several kinds according to their repair patterns and check if there is any bug in non-Mockito projects possessing the same repair pattern for each kind. Further, for each kind, we explore whether there are bugs in non-Mockito projects that have been successfully repaired before.

**RQ3: What are the performances of state-of-the-art APR techniques on Mockito?** This question is the focus of our research and is directly related to our conclusion. Generally, APR tools can be classified into two categories, i.e., search-based and semantics-based. For evaluating the performances of state-of-the-art tools on Mockito, we select two of the latest search-based tools, SimFix and CapGen, for performing the experiment and we also add the experimental results from another study [34], which presents the evaluation results on Mockito of an semantic-based tool, Nopol [35], into our analysis. We choose this technique report for analysis mainly because 1) Nopol is a representative semantic-based APR tool; and 2) we have no access to reproduce other semantic-based tools such as JFix at this time. (For details, see Section 5).

**RQ4: Is there any unique repair action for fixing Mockito bugs?** Repair pattern is an abstract concept and is composed of several different kinds of repair actions [33]. This question starts from a finer-grained granularity by identifying and analyzing the unique repair actions included in Mockito bugs. Our target is to provide practical guidance for future research.

## 4. RESULTS AND ANALYSIS

In this section, we present the results and answers to our four research questions.

## 4.1 RQ1: The Complexity of Patches in Mockito

To characterize patches, a study of Linux Kernel patching process [36] measures *locality of patches* through three indicators (i.e., files, hunks, and lines). Another previous study [11] annotates the Defects4J bugs with patch size and number of modified files to compute the complexity. A more recent study [33] has performed detailed analysis of the patch characteristics in Defects4J. They further select six indicators for analyzing patch features (patch size, number of chunks, number of modified files, number of modified methods, spreading of chunks, and number of modified classes). In our study, we select the former four features for evaluating the patch complexity. We do not take the latter two into considerations since 1) spreading of chunks is not directly related to patch complexity, e.g., if two single lines of addition appear at the beginning and end of a class, it can spread more than three chunks of modification in the middle of this class, nevertheless, it is not obvious which situation is more difficult to be repaired; 2) number of modified classes is highly related to the number of modified files as shown in [33], making this indicator redundant. Along their ideas, we give our own explanations as follows.

It is widely known that there are three types of code changes: addition, deletion, and modification. Addition and deletion appear as lines of codes are added or deleted consecutively or separately in source code. Modification appears as sequences of removed lines are straight followed by added lines or vice-versa. The patch size is the sum of the number of lines of these three types of code change in the patch. In previous study [3], authors used patch size as an indicator when evaluating the repair results by stating that the less the patch size is, the more manageable the patch is. Thus, this indicator is a key point to the patch complexity.

Composed by the combination of addition, deletion, and modification of lines, a chunk is a sequence of continuous changes in a file. The number of chunks of a patch can provide insights on how a patch is spread through the source code and further give information about how complex the patch is: the more chunks means the more buggy points in the program, and thus the more complexly for fixing this bug. Patches with more chunks are more complex in logical structure than patches with fewer chunks. Several empirical studies [11, 14] have proved this perspective.

Similarly, the number of modified files and the number of modified methods are also two important indicators. The larger they are, the more program elements are involved in the patch, and the more complex the patch is.

The previous study [33] has figured out the statistics of these indicators of bugs from Defects4J using the same standard as we mentioned and the data is publicly available in their online appendix[2]. We counted them and made the distribution of the patches of each project about these four features as the box plots in Figure 1. Note that in each subgraph, the bar displayed at the rightmost is the distribution of this feature for the patches of non-Mockito projects (five projects in Defects4J except Mockito).

1) *Patch size:* In Mockito, the average value of patch sizes is around 7 (7.08, accurately) while the value for other projects is 6.7. The median value is 4 and 25% of the patches change less than two lines, both these values the same as that of the non-Mockito projects.
2) *Number of chunks:* In Mockito, the average and median numbers are 3.5 and 3, both larger than that of the non-Mockito projects which are 2.6 and 2, respectively. The maximum of this feature is 20 and it occurs three times (one in Mockito, one in Lang, and one in Time).
3) *Number of modified files:* In Defects4J, 92.41% of the patches modify only one file [33]. Thus, the box in the figure becomes a line with the value of 1. Nevertheless, the average value of Mockito is 1.2, still slightly larger than that of the non-Mockito projects which is 1.08.
4) *Number of modified methods:* Although the first quartiles of Mockito and non-Mockito projects are both 1, the average of Mockito is 2.3, larger than that of non-Mockito projects (1.4). This time, the maximum of this feature is 20 and it only occurs once in Mockito-6.

From the results, we can see that the values of patch characteristics of Mockito are slightly larger than that of non-Mockito projects with respect to the four features. We further perform the significant difference test to check if the differences are significant. Our assumption is as follows:

**$H_0$: The complexities of Mockito's patches are distributed the same as the patches of non-Mockito projects on four indicators.**

We use the method named *Kolmogorov-Smirnov test* introduced in [37] to perform this test. The results are shown in Table 3. All the p-values are larger than 0.1, meaning that it is not significant at the 10% significance level, i.e., supporting the null hypothesis. Thus, the Mockito patch features are subject to the same distribution as the patch features of non-Mockito projects.

> **RQ1: Are patches of Mockito bugs more complex than that of non-Mockito bugs in Defects4J?**
>
> *Findings:* From four aspects (patch size, number of chunks, number of modified files, and number of modified methods), the averages of Mockito are slightly larger than that of non-Mockito projects. However, these differences are not significant at the 10% significance level according to a test. Thus, **Mockito bugs are not more significantly complex for repairing than bugs from non-Mockito projects**.

---

[2] http://program-repair.org/defects4j-dissection/

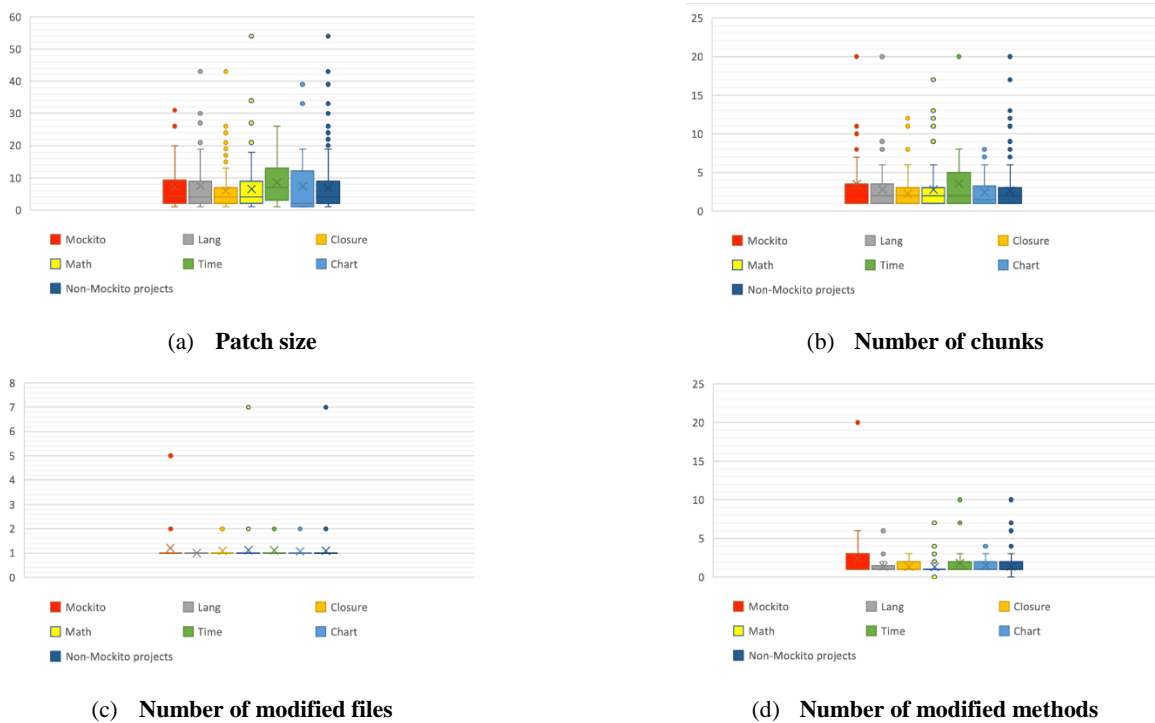

(a) **Patch size**  (b) **Number of chunks**

(c) **Number of modified files**  (d) **Number of modified methods**

**Figure 1. Features distribution map of the Defects4J patches**

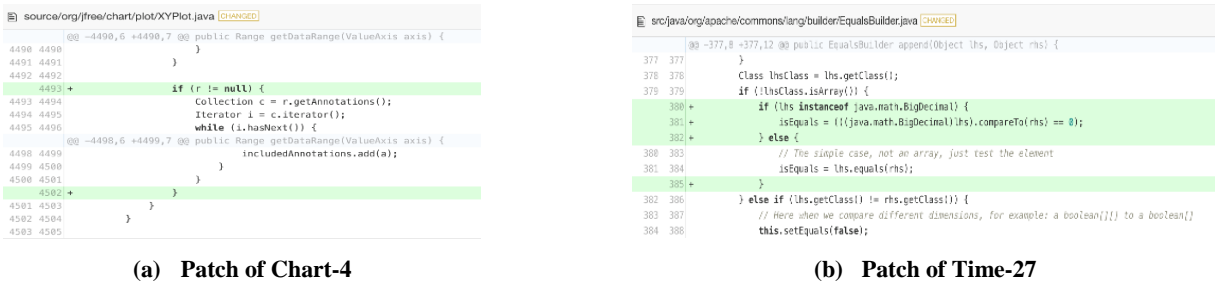

(a) **Patch of Chart-4**  (b) **Patch of Time-27**

**Figure 2. Two instances of the repair pattern: Wraps-with**

**Table 3. P-values of the four indicators**

| Indicators | P-values |
| --- | --- |
| Patch size | 0.998674 |
| Number of chunks | 0.159664 |
| Number of modified files | 0.999202 |
| Number of modified methods | 0.126142 |

### 4.2 RQ2: The Repair Patterns in Mockito

A repair action is a code operation (i.e., addition, removal, and modification) over a code element (e.g., conditional statement and method call) [33] and can be defined as a pair (<operation kind>, <element type>). Thus, repair action is often named in this form "element type + operation kind", e.g., *method call addition* as we will show in Section 4.4.

Repair pattern is a recurrent abstract structure in patches [32]. In the study [33], the repair patterns are established based on a Thematic Analysis (TA) process including identifying initial repair actions, combining repair actions re-appearing over many patches, and naming the themes. The differences between action and pattern can be explained through Figure 2. The repair actions are *if branch addition* and *if-else branches addition* for the former (Figure 2(a)) and the latter (Figure 2(b)), respectively, while both patches belong to the repair pattern named *Wraps-with* which represents the wrapping of an existing code with a conditional branch. This case interprets that the repair pattern is a collection of repair actions as we have introduced in Section 1.

Pioneered by Pattern-based Automatic program Repair (PAR) [38], there is a trend in APR of utilizing existent repair patterns for guiding the repair process [8, 25, 29]. Researchers have summarized more and more repair patterns with finer-grained granularity, from AST statement level [10, 39, 40] to expression level [12], and made great progress in using these patterns for fixing bugs. In this question, we aim to check whether the repair patterns summarized in previous studies are enough for bugs in Mockito and whether there exists any bug that has been fixed utilizing the same pattern with Mockito bugs.

The previous study [33] categorizes patches in Defects4J benchmark into nine repair patterns. We statistic the repair patterns of bugs in Mockito from their results and illustrate them in Table 4. Note that to obtain information from a more detailed perspective, the authors of [33] create several sub-categories under each category (e.g., to distinguish the wrapping structure, the variants for *Wraps-with* include *if*, *if-else*, and *method call* as is shown in Table 4). In the column "Repair patterns", we choose to demonstrate the sub-categories with the aim of investigating from a more detailed perspective. The column "Selected samples" shows some cases of bugs from non-Mockito projects which need the same repair pattern to fix. Due to the space limitation, we randomly select eight cases for each classification as the samples since the most common category (*Single line*) possesses 98 bugs (see the column "Total number of bugs"), making it impossible to list them all in the table. The selection is completely random without any bias and in these samples, the number of bugs of each project is calculated based on the proportion of its total bugs in this category. The column "Fixed bugs" at the rightmost lists the successfully repaired bugs in the column "Selected samples". The APR techniques we take into consideration when collecting these information come from ACS, Elixir, GPFL[3] [41], HDRepair [8], ssFix, JAID, CapGen, Nopol, and SimFix. In this paper, to refer to Defects4J bugs, we use a simple notation with project name followed by bug id, e.g., Math-5. Note that in this table, a bug can be classified into several categories (e.g., Mockito-19 is classified into *Conditional block addition* and *Addition with return statement*), that is due to the multiple edits in the patch: every code change chunk in the patch is analyzed and classified and if a patch contains several chunks, it may be classified into several categories.

**Table 4. Details of the repair patterns of Mockito bugs**

| Repair patterns | Total number of bugs | Mockito bugs# | Selected samples | Fixed bugs |
|---|---|---|---|---|
| Conditional block addition | 79 | 1,2,3,13,19,21, 23,33,36,37 | Lang-39,45 Time-1,2,13 Math-36,51,54 | Lang-39,45 |
| Addition with return statement | 77 | 4,9,11,18,19, 21,22,23 | Chart-14,15 Closure-33,60 Lang-5,40 Math-92,93 | Chart-14,15 Closure-33 Math-93 |
| Logic expression expansion | 48 | 34 | Chart-9 Closure-39,50 Lang-24,27,30 Math-32,37 | Chart-9 Lang-24,27 Math-32 |
| Logic expression reduction | 12 | 13 | Chart-5 Lang-15 Time-2 Closure-7,23,31,35,89 | Chart-5 Closure-31 |
| Logic expression modification | 49 | 17 | Chart-16 Closure-22,62,73 Lang-50 Math-47,94 Time-19 | Closure-62,73 Lang-50 Math-94 Time-19 |
| Wraps-with if statement | 24 | 14,15,16 | Chart-15 Closure-96 Lang-3,46 Math-28,95 Time-3,27 | Chart-15 Math-95 |
| Wraps-with if-else statement | 46 | 11,12,17, 24,29,38 | Closure-23,56,111 Lang-28,33 Math-47,97 Time-12 | Lang-33 Time-12 |
| Wraps-with method call | 14 | 4,14,28 | Chart-12,13 Closure-16 Time-8 Math-23,26,35,105 | Chart-12,13 Math-35,105 |
| Wrong reference variable | 42 | 6,20 | Chart-8,12 Closure-54 Time-4 Lang-4,60 Math-21,64,98 | Chart-8,12 Lang-60 Time-4 Math-98 |
| Wrong reference method | 31 | 6,32,35 | Chart-13 Closure-4,45,109 Lang-26 Math-58,75 Time-26 | Chart-13 Lang-26 Math-58,75 |
| Missing null check | 25 | 4,23,29,38 | Chart-15,25 Closure-20,30 Lang-32,33 Math-32 Time-2 | Chart-15,25 Lang-33 Math-32 |
| Missing non-null check | 32 | 33 | Chart-25 Closure-17,22,76,98 Lang-12 Math-20 Time-21 | Chart-25 Math-20 |
| Constant change | 19 | 26 | Closure-14,40,70 Lang-19 Math-22,104 Time-8,10 | Closure-14,40,70 Math-22,104 |
| Copy/Paste | 48 | 4,6,35 | Chart-7 Closure-4,6 Time-12 Lang-30,62 Math-37,76 | Chart-7 Time-12 |
| Single line | 98 | 5,7,8,24,26, 28,29,34,38 | Chart-1,24 Closure-62,67 Lang-51 Math-32,94 Time-16 | Chart-1,24 Closure-62 Lang-51 Math-32,94 |
| Not classified | 22 | 10,25,27,30,31 | Chart-3 Closure-25 Lang-35 Math-8,61,90 Time-7,22 | Chart-3 Lang-35 Time-7 Math-8,61,90 |

From the results, bugs in Mockito contain up to 15 kinds of repair patterns with 5 of them (Mockito-10, 25, 27, 30, 31) being not classified. *Conditional block addition* is the most popular pattern among this project, including totally ten bugs. Several patterns like *Logic expression expansion* and *Constant change* contain only one instance and thus are not very popular in this project. For each classification, at least two cases from samples have been successfully repaired in the past. At the most, six of the eight instances have been fixed before (*Single line* and *Not classified*). Even the bugs which are not classified can be fixed by current tools. According to the aforementioned analysis, Mockito bugs do not contain unique repair pattern and state-of-the-art APR techniques have achieved success on fixing bugs with the same repair patterns.

---

[3] In their study, authors do not name their technique. We name it GPFL for representing GenProg Fault Localization and being accordance with the study [32].

> **RQ2: Are state-of-the-art techniques capable to fix non-Mockito bugs whose repairs require the same repair patterns as Mockito bugs?**
>
> *Findings:* Most of the repair patterns of Mockito bugs (33/38) can be classified into existing categories. In each category, a number of bugs in non-Mockito projects have been successfully fixed by existing techniques. Thus, the **state-of-the-art techniques are capable of fixing bugs from non-Mockito projects whose repairs require the same repair patterns as Mockito bugs.**

### 4.3 RQ3: The Performances of State-of-the-art APR Techniques on Mockito

APR techniques can be generally divided into two categories (i.e., search-based and semantic-based). Search-based repair methods (also known as generate-and-validate methods [8]) search within a huge population of candidate patches generated by applying mutation operators to a predefined fault space determined by Fault Location (FL) techniques and are widely recognized to be able to fix a wide range of bugs [26, 41]. Semantics-based repair methodology, on the contrary, utilizes semantic information generated by symbolic execution and constraint solving to synthesize patches. In this section, we aim to check the performances of state-of-the-art APR tools on Mockito. To this end, we select two latest and open access search-based tools (SimFix and CapGen) for evaluating them on Mockito and we also take a reproduced study [34] about the semantic-based tool,

Nopol, into consideration when analyzing. All the execution logs of our experiments can be found in the link[4].

### 4.3.1 SimFix

SimFix is one of the latest search-based APR tools and has been evaluated on the non-Mockito projects in Defects4J [29]. The key novelty of SimFix is that it that takes the intersection of existing patches and source code into consideration to reduce search space. We first got the fault space information of suspicious buggy statements for each bug through the Ochiai [42] algorithm implemented in GZoltar version 1.6.0 [43] and then reran SimFix to evaluate its performances on the Mockito bugs. All the experiments were conducted on a 64-bit Linux virtual machine with Ubuntu 15.10 operating system and 2GB RAM. The execution results of SimFix can be three types: *success* when a patch passes all test cases, *failed* when all the suspicious statements are executed but still no patch is found, and *timeout* when the execution exceeds the pre-defined time. The experiment results are illustrated in Table 5.

As is shown, none of Mockito bugs can be fixed by SimFix. From the execution logs, ten of these bugs failed due to timeout and the rest 28 bugs were due to incapable of finding a valid patch at all suspicious locations. The execution time ranges from 61 minutes to 335 minutes and bugs with *timeout* result mostly ran around 5 hours except for Mockito-26. The reason for this outlier is that when the buggy program fails N test cases, SimFix arranges 300/N minutes to find an edit that can pass this test case. In this bug, there are totally four failing test cases and SimFix quickly finds correct edits for three of them but fails for the last one. Thus, the total execution time is much less compared with other bugs whose results are *timeout*. Generally, Automated program repair contains three steps: fault localization, patch generation, and patch validation. APR tools mainly differ at how to generate candidate patches while location techniques are integrated in the tools. We further checked if the unideal performances are due to deficient fault space generated by existing FL techniques by referring to human-written patches of each bug and checking if all the edit points were figured out by the Ochiai formula. Concretely, we consider the ground truth of developer patches and list the locations that have been modified as the faulty locations. We list the information in the rightmost column named "Fault space". "All" denotes all the edit points in the patches are listed in suspicious statements list. "Partial" denotes some of the edit points are included but some are not. It is specially for multi-edit bugs. "None" denotes none of the edit points are figured out by the location algorithm. It refers to both single-edit bugs and multi-edit bugs. Data in parentheses shaped like A/B format denotes the number of edit points calculated out by Ochiai algorithm (A) and the total number of edit points in the human-written patch of this bug (B). The Ochiai algorithm accurately figured out all the suspicious locations for 65.8% bugs (25/38) and only 5 bugs' error information were not calculated at all. In general, 78.5% of the buggy points (73/93) were presented to SimFix. We then manually added all lost edit points information to the corresponding bugs whose original location information is "None" or "Partial" and reconducted these 13 experiments. It is called *Line_Assumption* in a recent study [13] since we assume that the faulty code lines are known. A previous study [44] has shown that SimFix possesses the ability to repair multi-edit bugs. Thus, our intuition is that this time, with sufficient location information, SimFix at least can repair some bugs. However, results show that it still could not figure out any bug.

---

[4] [link redacted for double-blind review]

**Table 5. Experiment results of SimFix**

| Bug ID | Execution time (m) | Execution results | Fault space |
|---|---|---|---|
| 1 | 335 | Timeout | All(1/1) |
| 2 | 304 | Timeout | All(1/1) |
| 3 | 308 | Timeout | All(2/2) |
| 4 | 178 | Failed | None(0/3) |
| 5 | 62 | Failed | None(0/1) |
| 6 | 307 | Timeout | Partial(14/20) |
| 7 | 119 | Failed | All(1/1) |
| 8 | 154 | Failed | All(1/1) |
| 9 | 303 | Timeout | None(0/1) |
| 10 | 185 | Failed | Partial(4/5) |
| 11 | 190 | Failed | Partial(1/2) |
| 12 | 305 | Timeout | All(1/1) |
| 13 | 112 | Failed | All(1/1) |
| 14 | 104 | Failed | All(2/2) |
| 15 | 142 | Failed | All(1/1) |
| 16 | 125 | Failed | Partial(3/4) |
| 17 | 61 | Failed | Partial(2/3) |
| 18 | 271 | Failed | All(1/1) |
| 19 | 271 | Failed | Partial(7/8) |
| 20 | 296 | Failed | Partial(1/2) |
| 21 | 221 | Failed | All(4/4) |
| 22 | 90 | Failed | All(1/1) |
| 23 | 303 | Timeout | All(6/6) |
| 24 | 225 | Failed | All(1/1) |
| 25 | 113 | Failed | All(4/4) |
| 26 | 85 | Timeout | All(1/1) |
| 27 | 86 | Failed | All(1/1) |
| 28 | 79 | Failed | All(1/1) |
| 29 | 100 | Failed | All(1/1) |
| 30 | 118 | Failed | Partial(1/2) |
| 31 | 120 | Failed | All(1/1) |
| 32 | 112 | Failed | All(1/1) |
| 33 | 297 | Timeout | All(1/1) |
| 34 | 229 | Failed | All(1/1) |
| 35 | 302 | Timeout | All(3/3) |
| 36 | 116 | Failed | None(0/1) |
| 37 | 111 | Failed | All(1/1) |
| 38 | 188 | Failed | None(0/1) |

The failure of SimFix is mainly caused by two reasons. First is the lack of rich context information. Take Mockito-37 as an example, the human-written patch of this bug adds an method invocation (*reporter.cannotCallRealMethodOnInterface()*) under a conditional branch. However, this method invocation does not occur in any other place in this project, making SimFix incapable of generating a correct patch. Second is the course-grained donor snippet identification. Given a potential faulty location, SimFix expands it into a faulty code snippet and then locates a set of similar code snippets as donors for repair. The size of the code snippet can be as large as 10 lines. Consequently, if the correct fixing ingredient is only in a single line, SimFix may overlook it and fail to repair. For example, in Mockito-26, human-written patch turns a parameter in a method invocation from 0 to 0D. The correct fixing ingredient is several lines before this statement but the objects which invocate the same function at these two locations are not the same, making variable name similarity (i.e., a metric which is used by SimFix to identify similar donor) rather low. Thus, SimFix considers these two code snippets as dissimilar, leading to the miss of possible correct patch of this bug. From the aforementioned analysis, enlarging the fixing ingredients search space to contain some existing patches from open source projects (like ssFix) and developing a finer-grained donor snippet identification method may advance the performance of SimFix.

SimFix cannot fix any bug even if all the edit points are exposed. Thus, it performs poorly on Mockito bugs.

*4.3.2 CapGen*

CapGen is another state-of-the-art search-based APR tool that utilizes context information to prioritize patches. Empirically, its precision can reach 84% on the four projects of Defects4J [26]. We reran CapGen on Mockito bugs under the same environment which is consistent with its original execution environment in [26]. Note that the original testing framework in CapGen is GZoltar version 0.1.1 [26] whose localization capability is worse than that of the testing framework of SimFix (GZoltar version 1.6.0) [13]. Thus, this time, we changed the testing framework to version 1.6.0 with the hope that more bugs may be fixed when more accurate fault localization is provided by GZoltar 1.6.0 [13]. We still chose Ochiai algorithm for computing suspicious scores. The experiment results are illustrated in Table 6. As for execution time, CapGen finds out patches within 20 minutes for all bugs which is much quicker than SimFix. The columns "#Gen", "#Pla", and "#Cor" denote the numbers of generated patches, plausible patches, and correct patches for each bug, respectively. Totally, CapGen generates 5,212 patches for Mockito bugs, none of which is plausible (i.e., pass both the failing test cases and the regression test cases). Thus, it generates no correct patch. The fault space information is consistent with that of SimFix since both the testing framework and the FL ranking metric we exploit are the same. Still, we augmented the fault space with all buggy locations (i.e., inferred from the bug fixes) and reran the experiments for the 13 bugs whose buggy locations are not covered by the original fault space generated by the Ochiai algorithm. However, none of these bugs were fixed neither.

CapGen fails mainly because: 1) it performs a single mutation to generate patches, making it impossible to fix bugs which need operations at several different places (i.e., multi-location bugs) and this kind of bug occupies a large proportion in Mockito project (65.8%, 25/38); and 2) it usually cannot find the correct fixing ingredients since its searching scope is smaller compared with SimFix (*the single file* **vs.** *the whole project*). For example, in Mockito-27, human-written patch replaces a parameter of an object instantiation with another one (*oldMockHandler.getMockSettings()*) under the class *MockUtil*. However, the correct fixing ingredient is in another file named *MockHandler*, making it impossible for being extracted by CapGen. This point is easy to understand because the analysis in Section 4.3.1 has indicated that there may be no fixing ingredient even if the search space is the whole project. The above analysis shows the limitations discussed by the authors in [26] need to be overcome in the future. For the case of Mockito 5, which is likely to be successfully repaired by CapGen (both mutation operator and fixing ingredient exist), it requires to change the exception type *org.mockito.exceptions.verification.junit.ArgumentsAreDifferent* into *AssertionError*. Recall that CapGen needs to confirm that the type of a fixing ingredient match that of the target node before taking replacement operation to ensure the mutated program is compatible. However, it fails to infer that *AssertionError* is a compatible type of *ArgumentsAreDifferent*, and thus neglects this fixing ingredient consequently, leading to the failure of repairing this bug. This indicates that the context analysis ability of CapGen needs to be improved.

**Table 6. Experiment results of CapGen**

| Bug ID | Execution time (m) | #Gen | #Pla | #Cor | Fault space |
|---|---|---|---|---|---|
| 1 | 20 | 78 | 0 | -- | All(1/1) |
| 2 | 9 | 70 | 0 | -- | All(1/1) |
| 3 | 17 | 49 | 0 | -- | All(2/2) |
| 4 | 15 | 73 | 0 | -- | None(0/3) |
| 5 | 5 | 33 | 0 | -- | None(0/1) |
| 6 | 16 | 51 | 0 | -- | Partial(14/20) |
| 7 | 17 | 528 | 0 | -- | All(1/1) |
| 8 | 6 | 342 | 0 | -- | All(1/1) |
| 9 | 15 | 87 | 0 | -- | None(0/1) |
| 10 | 13 | 100 | 0 | -- | Partial(4/5) |
| 11 | 11 | 86 | 0 | -- | Partial(1/2) |
| 12 | 8 | 69 | 0 | -- | All(1/1) |
| 13 | 11 | 96 | 0 | -- | All(1/1) |
| 14 | 11 | 84 | 0 | -- | All(2/2) |
| 15 | 10 | 56 | 0 | -- | All(1/1) |
| 16 | 8 | 137 | 0 | -- | Partial(3/4) |
| 17 | 6 | 115 | 0 | -- | Partial(2/3) |
| 18 | 7 | 86 | 0 | -- | All(1/1) |
| 19 | 16 | 56 | 0 | -- | Partial(7/8) |
| 20 | 15 | 57 | 0 | -- | Partial(1/2) |
| 21 | 9 | 102 | 0 | -- | All(4/4) |
| 22 | 12 | 246 | 0 | -- | All(1/1) |
| 23 | 13 | 149 | 0 | -- | All(6/6) |
| 24 | 13 | 156 | 0 | -- | All(1/1) |
| 25 | 19 | 128 | 0 | -- | All(4/4) |
| 26 | 16 | 338 | 0 | -- | All(1/1) |
| 27 | 11 | 70 | 0 | -- | All(1/1) |
| 28 | 8 | 74 | 0 | -- | All(1/1) |
| 29 | 13 | 287 | 0 | -- | All(1/1) |
| 30 | 13 | 100 | 0 | -- | Partial(1/2) |
| 31 | 7 | 161 | 0 | -- | All(1/1) |
| 32 | 14 | 100 | 0 | -- | All(1/1) |
| 33 | 13 | 90 | 0 | -- | All(1/1) |
| 34 | 14 | 385 | 0 | -- | All(1/1) |
| 35 | 10 | 91 | 0 | -- | All(3/3) |
| 36 | 17 | 100 | 0 | -- | None(0/1) |
| 37 | 17 | 100 | 0 | -- | All(1/1) |
| 38 | 12 | 282 | 0 | -- | None(0/1) |
| Total | -- | 5,212 | 0 | -- | -- |

CapGen also performs poorly on Mockito bugs, fixing no bug under the assumption that all the buggy locations are known.

*4.3.3 Nopol*

Nopol is a semantic-based program repair tool utilizing angelic values and a Satisfiability Modulo Theory (SMT) solver for synthesizing conditional expressions [35]. A previous study has evaluated its performance on all the projects in Defects4J version 1.1.0 [34], thus, we adopt their results for our analysis here. Table 7 illustrates the empirical results of evaluating Nopol on Defects4J bugs.

It is shown in the results that Mockito is the lowest in terms of both the number of generated patches (2) and the repair rate (5%). The execution time is 18 seconds for Mocito-29 and 192 seconds for Mockito-38. Nopol totally generates patches for 103 Defects4J bugs and reaches an average repair rate of 26%. Note that these patches only pass all the test cases and are not manually checked. As a result, the repair rate here is consistent to the recall in other literature [21, 26] (i.e., the percentage of bugs for which Nopol generates a patch among all the bugs). We then manually analyzed these two patches and considered a patch correct if it is the same or semantically equivalent to human-written one provided in Defects4J, which is widely adopted by previous studies [21, 26, 29, 45]. According to our manual analysis, the patch for Mockito-29 is semantically equivalent to human-written one since they both wrap a statement with a conditional statement. However, the patch for Mockito-38 is not correct since it changes code at a wrong place and the generated code shown in Figure 3 is partially redundant and to some extent unreadable with too many mathematic symbols. Thus, Nopol actually only fixes one bug in Mockito, achieving a poor performance of a repair rate at around 2.6%.

**Table 7. Experiment results of Nopol**

| Projects | #Patches | #Bugs | Repair Rate |
|---|---|---|---|
| Chart | 9 | 26 | 37% |
| Closure | 56 | 133 | 42% |
| Math | 24 | 106 | 22% |
| Lang | 4 | 65 | 6% |
| Time | 8 | 27 | 29% |
| Mockito | 2 | 38 | 5% |
| Total | 103 | 395 | 26% |

```
--- /tmp/mockito_38_Nopol/src/org/mockito/internal/matchers/Equals.java
+++ /tmp/mockito_38_Nopol/src/org/mockito/internal/matchers/Equals.java
@@ -19,3 +19,3 @@
    public boolean matches(Object actual) {
-       if (this.wanted == null) {
+       if (!((actual!=null) && (actual!=null))) {
            return actual == null;
```

**Figure 3. Patch of Mockito-38 generated by Nopol**

---

**RQ3: What are the performances of state-of-the-art APR tools on Mockito?**

*Findings:* Both search-based APR approaches and semantic-based APR approaches possess extremely low repair rate on Mockito bugs. Compared with its performance on other projects, Nopol have lower repair rate on this project and can only fix one bug in fact. SimFix and CapGen even cannot fix any bug in this project. Thus, **the performances of the state-of-the-art APR tools on Mockito are poor**.

---

**Discussion:** According to the findings of our RQ1-3, Mockito bugs are not more complex with respect to the four patch characteristics (i.e., patch size, number of chunks, number of modified files, and number of modified methods); the state-of-the-art APR tools have achieved success on bugs whose repairs require the same repair patterns with Mockito bugs; three representative APR tools (one is semantic-based and two are search-based) achieve poor performances on this project. These are the bases of our argument. It is unreasonable to exclude Mockito from the verification set since the experimental results show that existing APR techniques may be overfitting to non-Mockito projects. Therefore, we recommend that **Mockito should be considered when evaluating newly proposed APR techniques to avoid potential bias from the evaluation process and judge the progress in repairing ability**.

## 4.4 RQ4: The Unique Repair Action in Mockito Bugs

Due to the poor performances of the state-of-the-art APR tools on Mockito, in this section, we further study whether there exists unique repair action in Mockito bugs that is ignored by previous studies, aiming to provide new insights for APR techniques.

The previous study [33] has classified the repair actions of each bug in Defects4J into corresponding category according to its operation kind and element type. We manually observed each category and checked whether the patches of Mockito bugs contain unique actions compared with patches from non-Mockito projects under that category. This was done by the first author and checked by the third author to confirm the correctness and comprehensiveness of our conclusion.

We find that under the categories named *method call addition* and *method definition addition*, some of the Mockito patches need to import some packages which are not contained in the program originally. We illustrate this situation through the patches of Closure-120 and Mockito-15 shown in Figure 4. They are both under the category named *method call addition* which means there is a call addition in the patch. However, when the patch of Closure-120 calls *getSymbol* and *getScope* APIs, it does not need to import an extra package, as a comparison, the patch of Mockito-15 needs to import the package named *BeanPropertySetter* to finish its call since it needs the fixing ingredients from that package. The study [33] does not distinguish these patches and classify them into the same kind, however, this is obvious different and can attribute to the failure of state-of-the-art tools on these bugs: they only concentrate on utilizing fixing ingredients that has already included in the program and do not make efforts for importing external sources, while the original fixing ingredients are not enough. As this phenomenon has not been discussed by any previous study, we comply with the naming rule and call this repair action *external package addition*. In total, Defects4J contains 13 bugs belonging to this category (11 from Mockito and 2 from Math). The eleven bugs from Mockito are Mockito-2, 9, 10, 14, 15, 17, 19, 25, 31, 32, and 36, occupying 28.9% (11/38) of the total amount of bugs in this project which indicates that this repair action is critical to fix bugs in this project. Another two bugs with the same repair action are Math-12 and 61. This small proportion (13/395) is in line with the conclusion from previous study [46] that most bugs can be fixed by rearranging existing code but this type appears mostly in Mockito (11/13), making it a unique repair action for this project. Thus, this category deserves investigation when studying Mockito. We then checked that if the two bugs from Math have been repaired by other tools and found that ACS fixes Math-61 successfully by importing the package named *NotStrictlyPositiveException* into *PoissonDistributionImpl* class[5] and thus enlarging the search space for fixing ingredients. After connecting with its developers, we get the information that ACS imports external packages by directly copying the imported packages in the test code to the source code. This experience provides a practical solution for searching for fixing ingredients which are not included in the original program: importing the external packages from the test code associated with the source code. Importing a suitable package will contain the correct fixing ingredients without the *search space explosion* problem [47].

When the original fixing ingredients are not enough, *external package addition* is a practical repair action for enlarging the search space. Copying imported packages in the relevant test code to the corresponding source code is a practical way learned from successful experience.

---

[5] Due to space limitation, we do not show the patch in this paper. Please see at: https://github.com/Adobee/ACS/tree/master/patch.

(a) Human-written patch of Closure-120

(b) Human-written patch of Mockito-15

Figure 4. Two patches for comparison

> **RQ4: Is there any unique repair action for fixing Mockito bugs?**
>
> *Findings: External package addition*, which refers to importing external sources for fixing bugs, is a unique repair action in Mockito, occurring in 28.9% of Mockito bugs. Copying imported packages in the test code to the source code is a practical solution for enlarging the search space of fixing ingredients.

## 5. LIMITATIONS

The limitations of this study mainly exist in the execution environment of the experiments and the generalization of the results.

**Environment.** Our experiments were conducted under 2GB RAM while SimFix was evaluated under 8GB. Note that this approach is search-based type (i.e., generate-and-validate with lots of calculation). Thus, low calculation ability may reduce the performances of this tool although there is no study concentrating on the influence of hardware configuration on the performances of APR tools.

**Generalization.** The results may not apply to other APR tools since APR is a hot topic in Software Engineering (SE) and there are lots of tools being released in main software conferences and journals each year but we only analyzed three of them. However, our evaluating subjects are representative since SimFix possesses the highest recall (34/357, 9.52%) and CapGen possesses the highest precision (21/25, 84%) on Defects4J among the state-of-the-art APR tools. Nopol generates the most patches for Defects4J bugs among semantic-based tools and is the most widely used semantic-based tool in empirical studies [44, 45]. Therefore, it is reasonable to speculate that other approaches may demonstrate similar phenomena. Besides, we deprecated other tools for several reasons: HDRepair, SOFix, ProbabilisticModel, and SketchFix do not release the source code of their tools for reproducible experiments. JFix and Elixir announce to design a plugin for Eclipse but do not provide the download address (the download link provided in JFix's homepage is invalid). ACS announces in its homepage[6] that it can no longer execute on new bugs due to the interface change in GitHub. ssFix needs a code search phase and the database is stored in a server in Brown university in the USA, making this process much slower for overseas users like us. JAID releases its source code but does not provide a *README.txt* file for guidance, making the reproducible experiment extremely time-consuming and error-prone. Thus, evaluating more APR tools on Mockito project can be future work.

---
[6] https://github.com/Adobee/ACS

## 6. RELATED WORK

During the years, developers have created several benchmarks for reproducible experiments on APR tools. The iBugs project [15] which contains 223 Java bugs with an exposing test case was initially created for fault localization. The software-artifact infrastructure repository (SIR) [16] can be considered as the first to provide a database of real bugs but most of its bugs are hand-seeded or obtained from mutation. ManyBugs and IntroClass benchmarks [18], containing 1,183 defects in 15 programs, are designed for C language and can be used for reproducing experiments on repair methods like GenProg. Then, based on the IntroClass benchmark, Durieux and Monperrus present IntroClassJava [17], which contains 297 small Java programs specified by Junit test cases. Recently, a multilingual program repair benchmark named QuixBugs [19] is designed with 40 programs in both Python and Java, each with a bug on one line. Although many benchmarks are created and applied, our evaluation object, Defects4J, is the most wildly-used one for Java language in recent studies [21-29].

There are also some experiments about evaluating previous tools on newly released benchmark. Martinez et al. [45] reimplement GenProg and Kali in Java language and evaluate them with Nopol on Defects4J benchmark. A recent technique report [34] evaluates GenProg on Defects4J version 1.1.0 including Mockito bugs and thus its experiment results are used by us for analysis. The ManyBugs and IntroClass [18] are used to present baseline experimental results for three previous repair methods (i.e., GenProg, AE [48], and RSRepair). Our study is the first to conduct experiments to present results for SimFix and CapGen on Mockito bugs.

There is a recent trend of analyzing the characteristics of existing patches to provide helpful information for automated repair process. The study [10] makes the first attempt to understand bug fixes in real world by concentrating on questions like fault localization, search space, and etc. Genesis [49] is the first repair system to automatically infer patch generation transforms from previous successful patches. Kim et al. [12] observe real-world patches from a expression-level granularity and provide nine insights for future APR researches. Wang et al. [14] study the curiousness that whether there are repeated bug fixes that change multiple program entities. The study [33] concludes the characteristics of human-written patches in Defects4J and the results are adopted by our analysis due to the validity of its study methodology.

Another empirical study [13] investigated bias caused by fault localization (FL) in APR and indicates that FL may be a reason for the failure of state-of-the-art methods to Mockito bugs. Our study concentrates on bias from the evaluation process and proves that two latest search-based APR techniques cannot fix Mockito bugs even if the faulty statement are known. Thus, our investigation can be considered as an extension of this study.

## 7. CONCLUSION

While Defects4J database is wildly used in evaluating the performances of APR tools, the Mockito project has not attracted enough attention. In this paper, we studied the importance of taking bugs from Mockito into consideration when evaluating APR techniques. We investigated the characteristics as well as the repair patterns of patches in Mockito. We conducted experiments on two of the latest search-based tools (SimFix and CapGen) on this project and took the experimental results of another semantic-based tool (Nopol) into consideration. Results show that state-of-the-art tools achieve poor performances on Mockito bugs. We thus reached the conclusion that indeed we should incorporate Mockito into the evaluation criteria. Further, we identified a unique repair action in Mockito bugs: *external package addition*. Importing the external packages from the test code associated with the source code is a feasible way to enlarge the search space.

## ACKNOWLEDGMENTS

Our thanks to the developers of Defects4J, SimFix, CapGen, and ACS for the kind help in our experiments.

## REFERENCES

[1] Britton T, Jeng L, Carver G, et al. Reversible debugging software[J]. Judge Bus. School, Univ. Cambridge, Cambridge, UK, Tech. Rep, 2013.
[2] Adamsen C Q , Anders Møller, Karim R , et al. Repairing Event Race Errors by Controlling Nondeterminism[C]// International Conference on Software Engineering. IEEE Press, 2017.
[3] Weimer W, Nguyen T V, Le Goues C, et al. Automatically finding patches using genetic programming[C]//Proceedings of the 31st International Conference on Software Engineering. IEEE Computer Society, 2009: 364-374.
[4] Qi Y, Mao X, Lei Y, et al. The strength of random search on automated program repair[C]//Proceedings of the 36th International Conference on Software Engineering. ACM, 2014: 254-265.
[5] Le X B D, Chu D H, Lo D, et al. S3: syntax-and semantic-guided repair synthesis via programming by examples[C]//Proceedings of the 2017 11th Joint Meeting on Foundations of Software Engineering. ACM, 2017: 593-604.
[6] Sidiroglou­douskos S , Lahtinen E , Long F , et al. Automatic Error Elimination by Horizontal Code Transfer Across Multiple Applications[J]. Acm Sigplan Notices, 2015, 50(6):43-54.
[7] Mechtaev S, Yi J, Roychoudhury A. Angelix: Scalable multiline program patch synthesis via symbolic analysis[C]//Proceedings of the 38th international conference on software engineering. ACM, 2016: 691-701.
[8] Le X B D, Lo D, Goues C L. History Driven Program Repair[C]// IEEE, International Conference on Software Analysis, Evolution, and Reengineering. IEEE, 2016:213-224.
[9] Nguyen H D T, Qi D, Roychoudhury A, et al. Semfix: Program repair via semantic analysis[C]//Software Engineering (ICSE), 2013 35th International Conference on. IEEE, 2013: 772-781.
[10] Zhong H, Su Z. An empirical study on real bug fixes[C]//Proceedings of the 37th International Conference on Software Engineering-Volume 1. IEEE Press, 2015: 913-923.
[11] Motwani M, Sankaranarayanan S, Just R, et al. Do automated program repair techniques repair hard and important bugs?[J]. Empirical Software Engineering, 2018, 23(5): 2901-2947.
[12] Liu K, Kim D, Koyuncu A, et al. A Closer Look at Real-World Patches[C]// In: Proceedings of IEEE International Conference on Software Maintenance and Evolution. IEEE, 2018: 304-315.
[13] Liu K, Koyuncu A, Bissyande T, et al. You cannot fix what you cannot find! An Investigation of fault localization bias in benchmarking automated program repair[C]// In: Proceedings of IEEE International Conference on Software Testing, Validation, and Verification (ICST). IEEE, 2019.
[14] Wang Y, Meng N, and Zhong H. An Empirical Study of Multi-Entity Changes in Real Bug Fixes[C]// In: Proceedings of IEEE International Conference on Software Maintenance and Evolution. IEEE, 2018: 316-327.
[15] Dallmeier V , Zimmermann T. Extraction of bug localization benchmarks from history[C]// IEEE/ACM International Conference on Automated Software Engineering. ACM, 2007.
[16] Do H , Elbaum S , Rothermel G. Supporting Controlled Experimentation with Testing Techniques: An Infrastructure and its Potential Impact[J]. Empirical Software Engineering, 2005, 10(4):405-435.
[17] Durieux T, Monperrus M. IntroClassJava: A Benchmark of 297 Small and Buggy Java Programs[D]. Universite Lille 1, 2016.
[18] Le Goues C, Holtschulte N, Smith E K, et al. The ManyBugs and IntroClass benchmarks for automated repair of C programs[J]. IEEE Transactions on Software Engineering, 2015, 41(12): 1236-1256.
[19] Lin D, Koppel J, Chen A, et al. QuixBugs: a multi-lingual program repair benchmark set based on the quixey challenge[C]// Proceedings Companion of the 2017 ACM SIGPLAN International Conference on Systems, Programming, Languages, and Applications: Software for Humanity. ACM, 2017: 55-56.
[20] Just R, Jalali D, Ernst M D. Defects4J: A database of existing faults to enable controlled testing studies for Java programs[C]// Proceedings of the 2014 International Symposium on Software Testing and Analysis. ACM, 2014: 437-440.
[21] Xiong Y, Wang J, Yan R, et al. Precise condition synthesis for program repair[C]//Proceedings of the 39th International Conference on Software Engineering. IEEE Press, 2017: 416-426.
[22] Xin Q, Reiss S P. Leveraging syntax-related code for automated program repair[C]//Proceedings of the 32nd IEEE/ACM International Conference on Automated Software Engineering. IEEE Press, 2017: 660-670.
[23] Chen L, Pei Y, Furia C A. Contract-based program repair without the contracts[C]//Automated Software Engineering (ASE), 2017 32nd IEEE/ACM International Conference on. IEEE, 2017: 637-647.
[24] Liu X, Zhong H. Mining stackoverflow for program repair[C]//2018 IEEE 25th International Conference on Software Analysis, Evolution and Reengineering (SANER). IEEE, 2018: 118-129.
[25] Soto M, Le Goues C. Using a probabilistic model to predict bug fixes[C]//2018 IEEE 25th International Conference on Software Analysis, Evolution and Reengineering (SANER). IEEE, 2018: 221-231.
[26] Wen M, Chen J, Wu R, et al. Context-Aware Patch Generation for Better Automated Program Repair[C]// Proceedings of the 40th International Conference on Software Engineering. ACM, 2018.
[27] Hua J, Zhang M, Wang K, et al. Towards practical program repair with on-demand candidate generation[C]//Proceedings of the 40th International Conference on Software Engineering. ACM, 2018: 12-23.
[28] Saha R K, Yoshida H, Prasad M R, et al. Elixir: an automated repair tool for Java programs[C]//Proceedings of the 40th International Conference on Software Engineering: Companion Proceeedings. ACM, 2018: 77-80.
[29] Jiang J, Xiong Y, Zhang H, et al. Shaping Program Repair Space with Existing Patches and Similar Code[C]// The International Symposium on Software Testing and Analysis. 2018.
[30] Le X B D, Chu D H, Lo D, et al. JFIX: semantics-based repair of Java programs via symbolic PathFinder[C]//Proceedings of the 26th ACM SIGSOFT International Symposium on Software Testing and Analysis. ACM, 2017: 376-379.
[31] Durieux T, Cornu B, Seinturier L, et al. Dynamic patch generation for null pointer exceptions using metaprogramming[C]//Software Analysis, Evolution and Reengineering (SANER), 2017 IEEE 24th International Conference on. IEEE, 2017: 349-358.
[32] Martinez M, Monperrus M. Mining software repair models for reasoning on the search space of automated program fixing[J]. Empirical Software Engineering, 2015, 20(1): 176-205.
[33] Sobreira V, Durieux T, Madeiral F, et al. Dissection of a bug dataset: Anatomy of 395 patches from Defects4J[C]//2018 IEEE 25th International Conference on Software Analysis, Evolution and Reengineering (SANER). IEEE, 2018: 130-140.
[34] Durieux T, Danglot B, Yu Z, et al. The patches of the nopol automatic repair system on the bugs of defects4j version 1.1.0[D]. Université Lille 1-Sciences et Technologies, 2017.
[35] Xuan J, Martinez M, Demarco F, et al. Nopol: Automatic repair of conditional statement bugs in java programs[J]. IEEE Transactions on Software Engineering, 2017, 43(1): 34-55.
[36] Koyuncu A, Bissyandé T F, Kim D, et al. Impact of tool support in patch construction[C]//Proceedings of the 26th ACM SIGSOFT International Symposium on Software Testing and Analysis. ACM, 2017: 237-248.
[37] Massey Jr F J. The Kolmogorov-Smirnov test for goodness of fit[J]. Journal of the American statistical Association, 1951, 46(253): 68-78.
[38] Kim D , Nam J , Song J , et al . Automatic patch generation learned from human-written patches[C]// 2013 35th International Conference on Software Engineering (ICSE). IEEE Computer Society, 2013.
[39] Martinez M, Monperrus M. Mining software repair models for reasoning on the search space of automated program fixing[J]. Empirical Software Engineering, 2015, 20(1): 176-205.
[40] Pan K, Kim S, Whitehead E J. Toward an understanding of bug fix patterns[J]. Empirical Software Engineering, 2009, 14(3): 286-315.
[41] Wen M, Chen J, Wu R, et al. An empirical analysis of the influence of fault space on search-based automated program repair[J]. arXiv preprint arXiv:1707.05172, 2017.
[42] Abreu R, Zoeteweij P, Van Gemund A J C. An evaluation of similarity coefficients for software fault localization[C]//Dependable Computing, 2006. PRDC'06. 12th Pacific Rim International Symposium on. IEEE, 2006: 39-46.
[43] Pearson S, Campos J, Just R, et al. Evaluating and improving fault localization[C]//Proceedings of the 39th International Conference on Software Engineering. IEEE Press, 2017: 609-620.
[44] Wang S, Mao X, Niu N, et al. Multi-Location Program Repair: Roads Ahead[J]. arXiv preprint arXiv:1810.12556, 2018.
[45] Martinez M, Durieux T, Sommerard R, et al. Automatic repair of real bugs in java: A large-scale experiment on the defects4j dataset[J]. Empirical Software Engineering, 2017, 22(4): 1936-1964.
[46] Martinez M, Weimer W, Monperrus M. Do the fix ingredients already exist? an empirical inquiry into the redundancy assumptions of program repair approaches[C]// Companion International Conference on Software Engineering. 2014.
[47] Claire Le Goues, Michael Dewey-Vogt, Stephanie Forrest, and Westley Weimer. 2012. A systematic study of automated program repair: Fixing 55 out of 105 bugs for 8 each. In ICSE'2012. IEEE, 3–13.
[48] Weimer W, Fry Z P, Forrest S. Leveraging program equivalence for adaptive program repair: Models and first results[C]//Automated Software Engineering (ASE), 2013 IEEE/ACM 28th International Conference on. IEEE, 2013: 356-366.

[49] Long F, Amidon P, Rinard M. Automatic inference of code transforms for patch generation[C]//Proceedings of the 2017 11th Joint Meeting on Foundations of Software Engineering. ACM, 2017: 727-739.